\documentclass[reprint,amsmath,amssymb,aps,prd,floatfix]{revtex4-2}
\usepackage{graphicx}

\begin{document}

\title{Topological signature of the quantum nature of gravity from the Pancharatnam phase in dual Stern-Gerlach interferometers}

\author {Samuel Moukouri}

\affiliation{Department of Physics, Ben-Gurion University of the Negev, Be'er Sheva 84105, Israel}
\thanks{Current Address: bomono LTD, 75 Haatsmaut, Ashdod 7745268, Israel}
\email{moukouri@bomono.tech}

\begin{abstract}
 Entanglement plays a central role in the fundamental testing and practical applications of quantum mechanics. Because entanglement is a feature unique to quantum systems, its observation provides evidence of quantumness. Hence, if  gravity can generate entanglement between quantum superpositions, this indicates that quantum amplitudes are field sources and that gravity is quantum. I study dual spin-one-half Stern-Gerlach interferometers (SGI) and show that the Pancharatnam phase is a tool that  qualitatively distinguishes semiclassical from quantum gravity. The semiclassical evolution is equivalent to that of two independent interferometers in an external gravitational field. In this case,  a phase jump was observed, as expected from the geodesic rule, which dictates the noncyclic evolution in the Bloch sphere. In contrast, in the quantum case, the quantum amplitudes are the sources of the gravitational field, inducing entanglement between the two interferometers, and the phase is continuous.
\end{abstract}

\maketitle
\section{Introduction} 
One of the distinctive features of quantum mechanical interactions is their ability to create entanglements between systems. Entangled systems are inherently non-local, which fundamentally distinguishes them from their local classical counterparts. Classical interactions cannot create entanglement between particles. The expression 'classical entanglement' found in the literature \cite{enk,azzini} refers to the impossibility of writing a single-particle wave function with different degrees of freedom as a product state. However, even if the single-particle wavefunction cannot be factorized into a product state of these degrees of freedom, physics remains local.  

Attempts to integrate general relativity (GR), albeit linearized GR, in the quantum mechanical framework began in the early years of quantum mechanics \cite{rosenfeld,bronstein,salomon}. However, these efforts aimed at incorporating the geometric formalism of GR into quantum mechanics have not reached completion. A different approach, which is more intuitive and closer to experiments, was suggested by Feynman. One of the experiments he devised was to use the Stern-Gerlach  interferometer (SGI) to detect a possible quantum signature of gravity \cite{feynman, zeh}. The Colella-Overhauser-Werner  experiment \cite{cow} yielded half of the proof, namely that quantum superpositions remain coherent when subjected to external gravitational fields.  This property is now routinely used in matter-wave interferometry. The essential question that remains unanswered is whether the quantum amplitudes in the SG apparatus can be sources of gravitational fields. If so, gravity can induce entanglement as shown by Bose et al.\,\cite{bose} who proposed testing gravity using dual spin-one-half SGIs \cite{SG,NatureSplit,Scully1987}. Hence, the emergence of entanglement is a direct consequence of considering the quantum amplitudes as sources of gravitational fields. Another tabletop experiment was proposed in Ref.\cite{marletto} simultaneously, Ref.\cite{bose,marletto} kicked off intense activity in tabletop experiments for quantum gravity \cite{huggett}. Although these proposed experiments still have to overcome technical challenges, they have the merit of defining the quantum gravity problem as a measurable laboratory experiment. Furthermore, the recent realization of complete SGI in cold atom systems has added impetus to the possibility of testing the quantum nature of GR using SGIs \cite{Margalit2021}.

\begin{figure*}[tbhp]
\centering
\includegraphics[width=\textwidth]{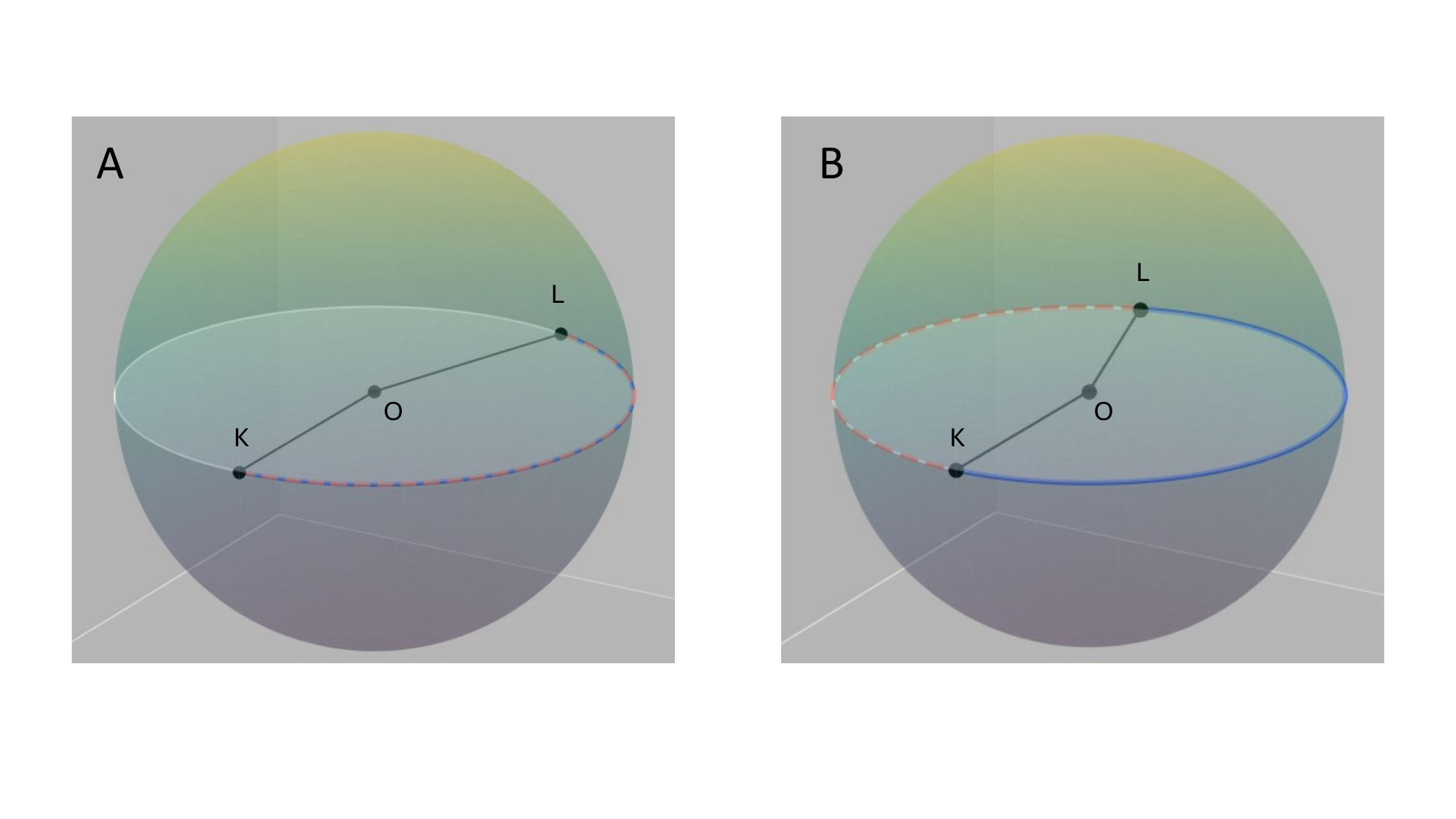}
\vspace{-2.cm}
\caption{Illustration of the geodesic rule for the Pancharatnam phase (which includes purely geometric and dynamical components): During the evolution from points K to L on the Bloch sphere, the geometric component of the Pancharatnam phase is given by minus half the spherical area enclosed by the closed curve formed by curve KL (blue line) and the shortest geodesic joining L to K. An orientation is given by the normal to the plane KOL. The geodesic rule is illustrated for a trajectory along the equatorial plane. The curve KL is shown by the blue line, and the shortest geodesic joining L to K by the dashed red line. (A) When $\widehat {\text{KOL}} < \pi$, the shortest geodesic joining L to K coincides with curve KL and the enclosed area on the Bloch sphere, and the Pancharatnam phase reduces to zero. (B) When $\widehat{\text{KOL}} > \pi$, curve KL and the shortest geodesic joining L to K form an equatorial circle, the enclosed spherical area is half of the Bloch sphere whose magnitude is $2\pi$ and the geometric phase is $-\pi$. There is a sudden change in the geodesic when $\widehat{\text{KOL}}$ crosses $\pi$ resulting in the jump of the Pancharatnam phase.}
\label{fig:geo-rule}
\end{figure*}

A central assumption underlying the Bose et al. proposal is that the observation of gravitational entanglement unambiguously signals the quantum nature of gravity, since classical interactions are held to be incapable of generating entanglement. This assumption has recently been challenged. Aziz and Howl \cite{aziz-howl} argued that even a purely classical gravitational field could, through virtual-matter processes, produce entanglement indirectly, thereby undermining the interpretational power of a positive experimental result. This claim was subsequently contested on theoretical grounds. Di\'osi \cite{diosi} demonstrated through exact, non-perturbative field-theoretic calculations that classical gravity does not, in fact, entangle quantized matter fields, and showed that the apparent entangling effect reported by Aziz and Howl arises from their perturbative treatment. Complementary analyses based on semiclassical and stochastic gravity models have reached the same conclusion: when the tidal gravitational field sourced by the quantum bodies remains classical, the composite state of the two masses is not entangled \cite{li-mondal}. Nevertheless, this debate reveals that a purely entanglement-based witness may be vulnerable to ambiguities in interpretation, particularly if experimental imperfections or alternative theoretical frameworks are invoked. It is therefore highly desirable to identify an observable that provides a qualitative, rather than merely quantitative, distinction between semiclassical and quantum gravity. The Pancharatnam phase, as we show in this work, offers precisely such a discriminator: it does not merely differ in magnitude between the two cases, but exhibits a fundamentally different topological character—a sharp discontinuous jump in the semiclassical regime versus a smooth, continuous evolution in the quantum regime. This structural difference persists even at low interferometric visibility, making it robust to many of the practical limitations that afflict entanglement-strength measurements.

Recent matter wave experiments with external magnetic gradients have shown that a single SGI can generate a noncyclic geometric phase,\cite{GeoPhaseJump}. The geometric phase \cite{Berry} arising from noncyclic transformations was first theoretically discussed by Samuel and Bhandari,\cite{samuel-bhandari}. The specific case of a two-level system was extensively explored in subsequent works by Bhandari, who pointed out the occurrence of acute phase jumps during evolution \cite{bhandari_1,bhandari_2}. When the system evolves from point K to point L on the Bloch sphere, the accumulated geometric phase is given by Pancharatnam’s theorem \cite{pancharatnam}. It is equal to minus one half of the spherical area on the Bloch sphere enclosed by trajectory KL and the shortest geodesic joining K and L. The phase jumps result from a sudden change of geodesic when the system crosses specific points of the Bloch sphere. This is illustrated in Fig.\ref{fig:geo-rule} for evolution around the equatorial plane. These phase jumps, called $SU(2)$ phase jumps by Bhandari, have been observed experimentally in various systems \,\cite{GeoPhaseJump,morinaga,vandijk,wagh,filipp,morinaga2,morinaga3}.

\begin{figure*}[tbhp]
\centering
\includegraphics[width=\textwidth]{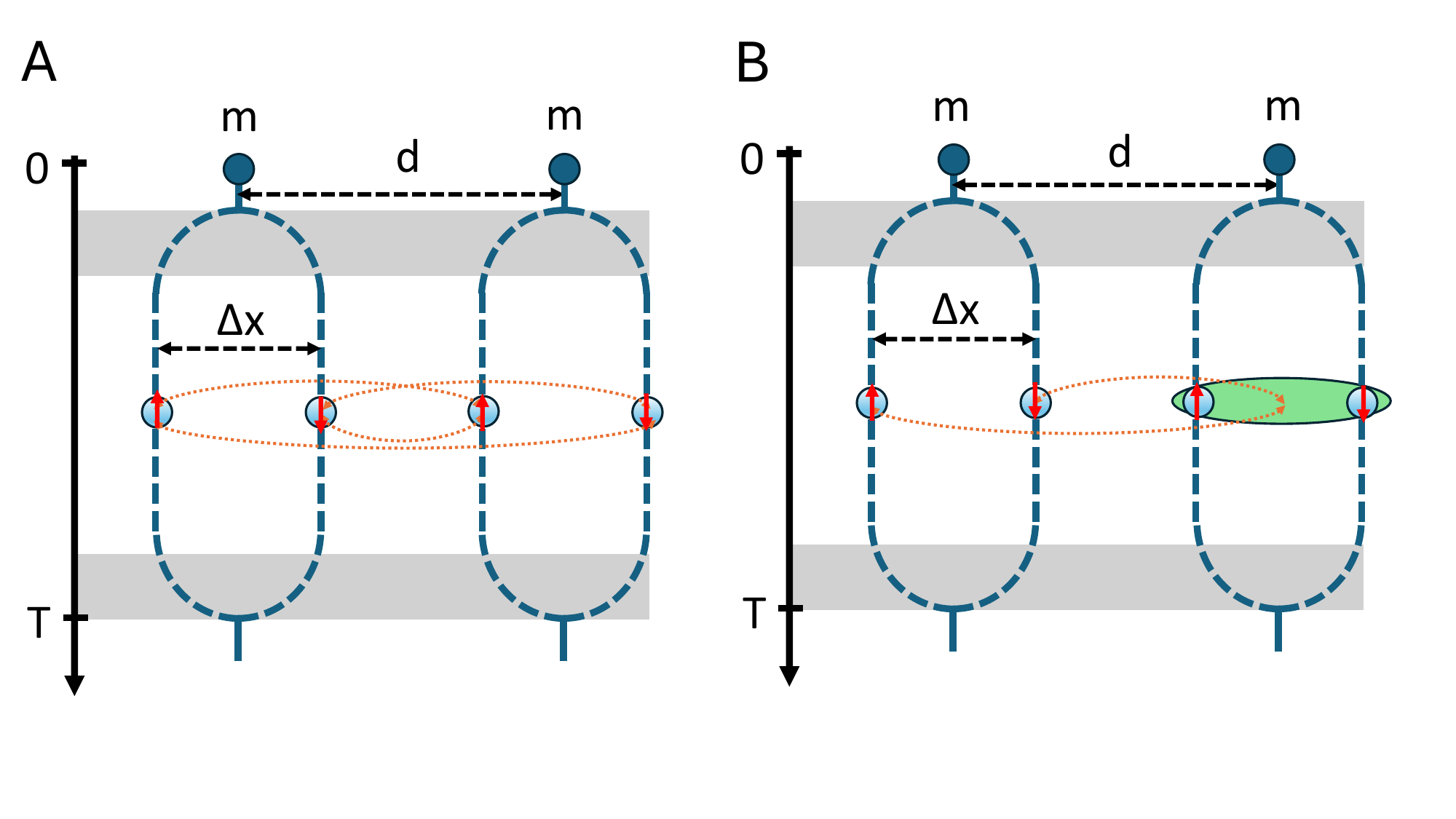}
\vspace{-2cm}
\caption{Sketch of the experiment proposed in Ref. \,\cite{bose}: two nanoparticles with equal masses $m$  and spins one-half are released from a trap at $t=0$. After a short time, an equal-weight superposition of each mass was created by an RF pulse of duration $\Delta t_{RF}$. The superposition of the nanoparticles can evolve under mutual gravity. They were then recombined by another RF pulse. The trap release time and $\Delta t_{RF}$ (grey area) are negligible compared to the free-fall time of the two superpositions. Here, I illustrate two possibilities. (A) Quantized gravity: quantum amplitudes are sources of field, and each superposition is sensitive to the gravitational field created by each component of the other superposition (note that the self-gravity term, which in principle should be included, does not induce a phase difference); (B) semiclassical gravity: the quantum amplitudes are not direct sources of gravitational field, and each superposition evolves under a single field created by the effective mass of the other superposition (green oval). In both cases, the interactions are indicated by dotted orange lines.}
\label{fig:sketch}
\end{figure*}

There are two alternative interpretations of the effects of gravity in dual SGIs. First, gravity is semiclassical, and the two SG interferometers evolve separately as two spin-one-half subsystems, each under the local gravitational field of the other. This situation is similar to that of  an SG interferometer in an external field as in Ref.\,\cite{GeoPhaseJump,FGBS,Margalit2015,QuCom,Amit2022}, the gravitational field acts as a magnetic field. For appropriate input parameters, the Pancharatnam phase  displays jumps \cite{GeoPhaseJump} because the  system is described by the $SU(2)$ symmetry as in Ref.\,\cite{bhandari_1}. Second, gravity is quantized, meaning that the superpositions are sources of the gravitational field, the two spin-one-half systems are entangled, and the system should be described by a larger subgroup of $SU(4)$, presumably $SO(5)$ because the two two-level systems are identical \cite{uskov}. As I will show in section \,\ref{sec:phase}, the Pancharatnam phase is radically different in the two situations, enabling the possibility of clearly distinguishing the quantum behavior of gravity from the semiclassical. It is important to stress that the proposed experiment will provide evidence of the superposition of quantum geometries but is non-relativistic; it is not able to validate or disprove theories of quantum gravity \cite{rovelli}. However, if masses on the order of Planck's mass are reached, discreteness of time, a prediction of loop quantum gravity, could be observed in this type of experiment \cite{rovelli_2}.

\section{Dual Stern-Gerlach interferometers} 
Sketch of the dual SGIs proposed in Ref. \,\cite{bose} are shown in Fig.\,\ref{fig:sketch}. Two spin-one-half nanoparticles of identical masses, $m$, are released from traps spatially separated by a distance $d$ at $t=0$; and then placed in superpositions of equal amplitude by an RF pulse of duration $\Delta t_{RF}$ (gray area), which is small compared to the interferometer time $T$ and will thus be neglected in the calculation of the phase accumulated during evolution. During $T$, the two superpositions evolve under mutual gravitational fields.  If gravity is quantum, as shown in Fig.\,\ref{fig:sketch}A, each amplitude in the superposition of a first interferometer is a source of  gravitational field. Hence, each branch of the second interferometer is influenced by the sources' fields at different locations. This leads to an entanglement between the two interferometers \cite{bose,marshman}.  However, if gravity is assumed to be semiclassical, each superposition creates an effective semiclassical field under which the other field evolves. In the illustration in Fig.\,\ref{fig:sketch}B the left superposition is only subjected to a field generated by the average mass of the two branches of the right superposition (green oval), and vice versa. The latter situation is similar to that of the usual studies of a spin-one-half SGI in a magnetic field \cite{GeoPhaseJump}. 

\begin{figure*}[tbhp]
\centering
\includegraphics[width=\textwidth]{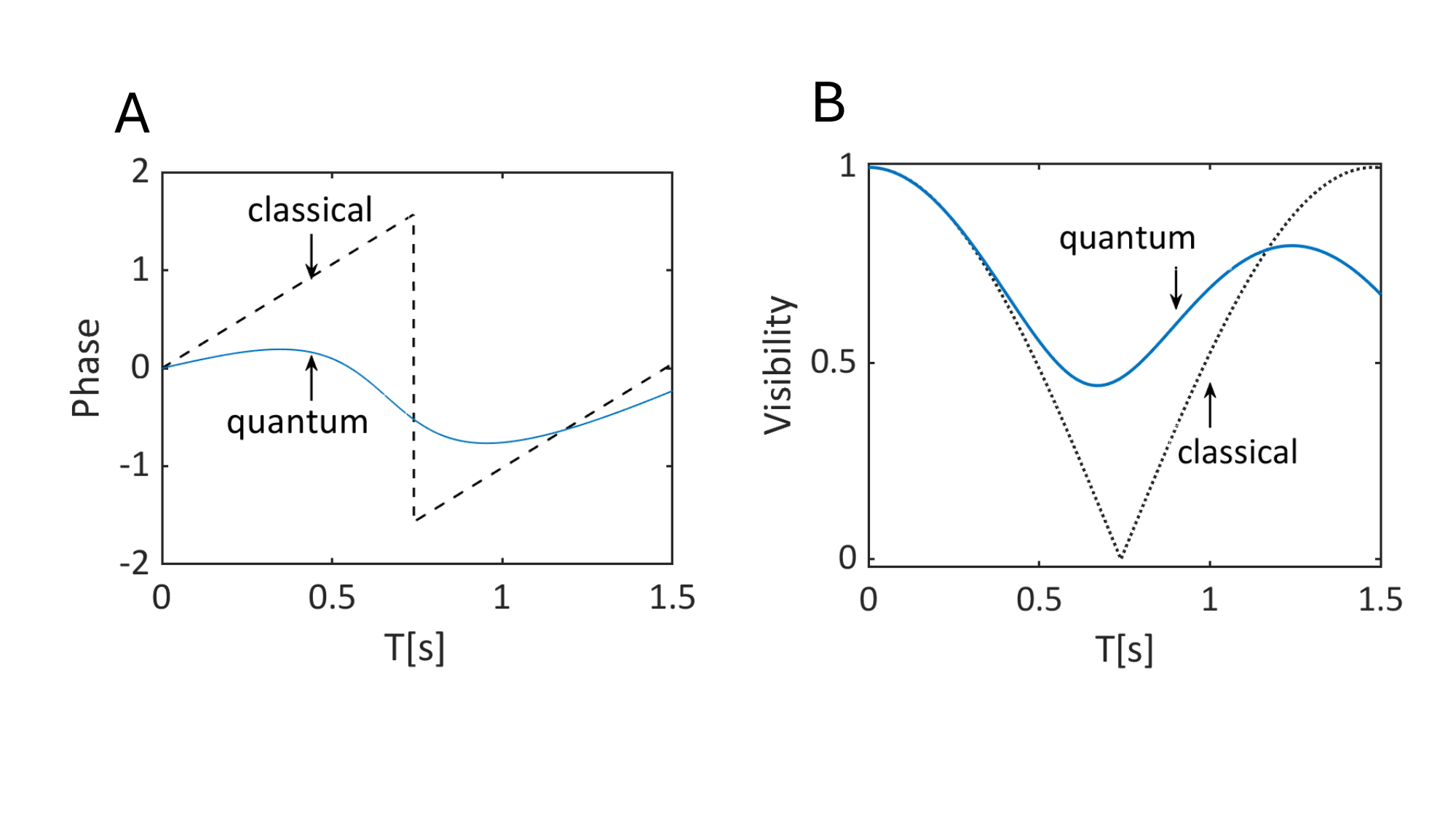}
\vspace{-2cm}
\caption{Interferometric phase and visibility: (A) The Pancharatnam phase in the dual SGI experiment proposed in Ref.\,\cite{bose}. The black dashed line corresponds to the semiclassical configuration discussed in Fig.\,\ref{fig:sketch}A. The continuous blue line corresponds to the quantum configuration of Fig.\,\ref{fig:sketch}B. (B) The visibility in the dual SGI experiment proposed in Ref.\,\cite{bose}. The black dashed line corresponds to the semiclassical configuration discussed in Fig.\,\ref{fig:sketch}A. The continuous blue line corresponds to the quantum configuration of Fig.\,\ref{fig:sketch}B.}
\label{fig:phase-vis}
\end{figure*}

\section{Pancharatnam phase}
\label{sec:phase} 

I apply the quantum kinematic theory of Ref.\,\cite{mukunda} to compute the Pancharatnam phase, by assuming semiclassical or quantum behavior of the dual SGI. The semiclassical and quantum phases, $\Phi_C$ and $\Phi_Q$ are respectively given by (see Methods),

\begin{equation}
\begin{aligned}
\Phi_C &= \arctan \frac{\sin \phi}{1+\cos \phi},\\
\Phi_Q &= \arctan\frac{\sin \frac{\phi_1+\phi_2}{2} \cos\frac{\phi_1-\phi_2}{2}}{1+\cos\frac{\phi_1+\phi_2}{2} \cos\frac{\phi_1-\phi_2}{2}},
\end{aligned}
\label{eq:phase}
\end{equation}

The essential difference between quantum and semiclassical calculations is that, in the quantum case, it is assumed that the amplitudes in the quantum superpositions are sources of the gravitational field. This is not obvious prior to a measurement. As a consequence, there is a factor $\xi=\cos[(\phi_1-\phi_2)/2]$ in Eq.\,\ref{eq:phase} which underscores the difference between semiclassical and quantum gravity. The difference in the positions of the gravitational sources in the superposition introduces a 'contrast' $\xi$ which suppresses the sharp phase jump of the semiclassical phase $\Phi_C$ as I show in Fig.\,\ref{fig:phase-vis}.

\section{semiclassical and quantum Phases} 
In Fig.\,\ref{fig:phase-vis}, I display $\Phi_C$ and $\Phi_Q$ as functions of the time of the interferometer. I chose parameters similar to those proposed in Ref.\,\cite{bose}: $m_0= 5 \times 10^{-14} kg,\,d=450\,\mu m,\, \Delta x=250\,\mu m,\,T=1.5\,s$. The mass, $m_0$, was chosen to achieve sufficiently large phase differences at $T=0.75$\,s, $\phi_1=-0.95$\,rad, and $\phi_2=6.26$\,rad.  In semiclassical gravity, $\Phi_C$ displays an acute phase jump of magnitude $\pi$, similar to conventional single-SG interferometers under magnetic gradients. Here, the gradients are provided by the effective gravitational field of the superposition (green oval) from one interferometer to the other. These phase jumps can be detected with high accuracy and can be used for precision amplification in metrology \cite{zhou24}. However, when the quantum amplitudes in each branch of the interferometer are considered sources of the gravitational field, the sharp singularity disappears, as seen in the plot of $\Phi_Q$ in Fig.\,\ref{fig:phase-vis}. The phase singularity displayed in the semiclassical case becomes an inflection point.

In Fig.\,\ref{fig:phase-vis}B, I show the visibility, $V=\left|\langle \Psi(0) | \Psi(t) \rangle \right|$, as a function of the interferometer time $T$. In the semiclassical system \cite{GeoPhaseJump} which is dictated by the geodesic rule, the visibility vanishes at the singularity point where the phase $\Phi_C$ jumps. This is illustrated in Fig.\,\ref{fig:phase-vis} where the visibility vanishes near $T=0.75$\,s at the position of the singularity. In contrast, in the absence of a singularity, the visibility is not expected to go to zero, it displays a non-zero minimum near $T=0.75$\,s.

 \begin{figure*}[tbhp]
\centering
\includegraphics[width=\textwidth]{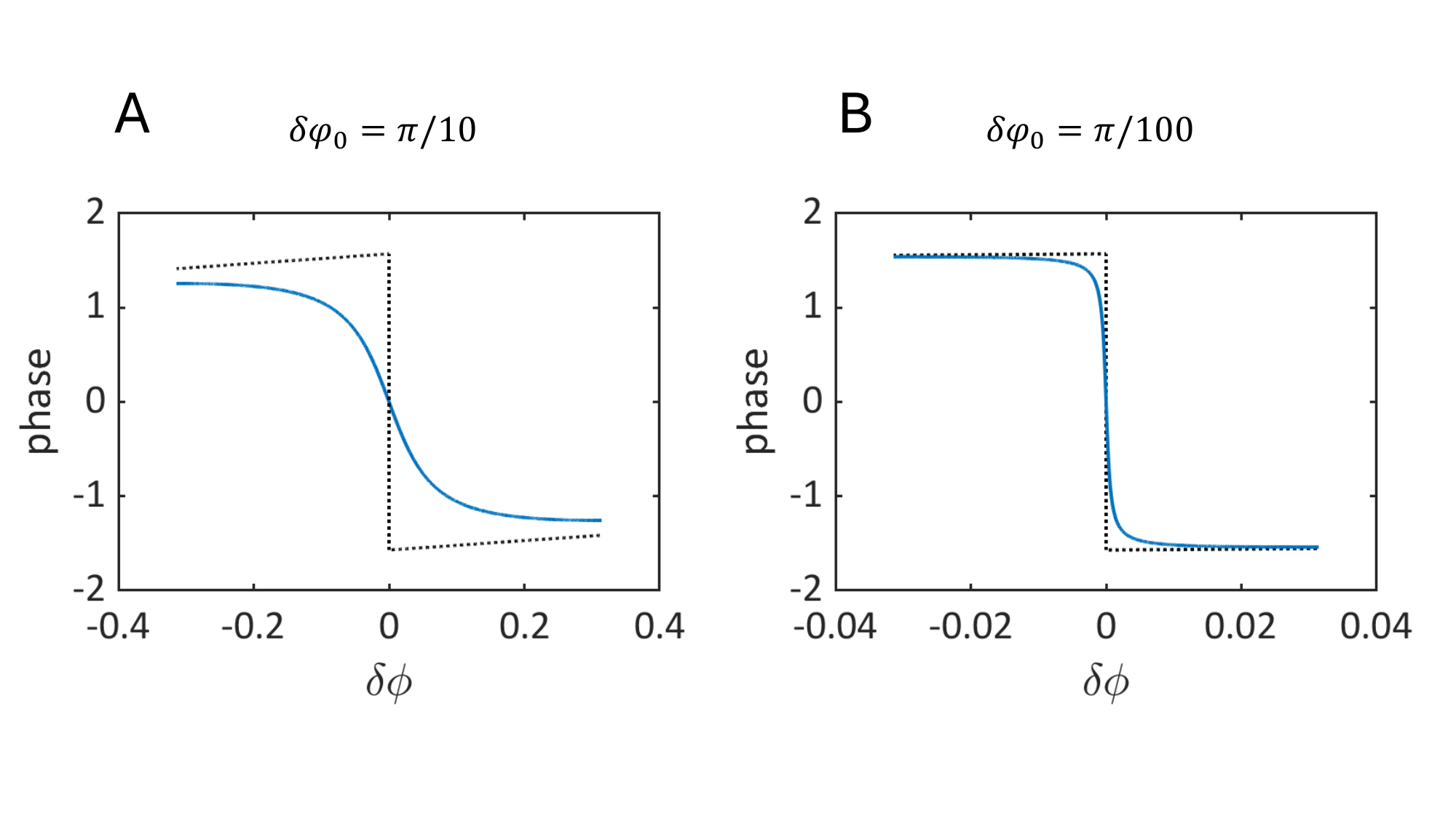}
\vspace{-2cm}
\caption{ The Pancharatnam phase in the dual SGI experiment proposed in Ref.\,\cite{bose} near the semiclassical singularity shown in Fig.\,\ref{fig:phase-vis}. The system is brought at $\delta \phi_0$ from the singularity by a magnetic gradient pulse and then let evolve under the gravitational interaction. The black dashed line corresponds to the semiclassical interferometer configuration discussed in Fig.\,\ref{fig:sketch}A. The continuous blue line corresponds to the quantum configuration of Fig.\,\ref{fig:sketch}B. (A) $\delta \phi_0=\pi/10$; (B) $\delta \phi_0=\pi/100$.}
\label{fig:phase-wpoint}
\end{figure*}

\section{Discussion and Outlook}
The Pancharatnam phase provides a means to qualitatively distinguish semiclassical from quantum gravity. This stems from the inherent difference in the symmetry of semiclassical and quantum description of the coupled SGI. The former is described by two independent interferometers in an external field while in the latter the two interferometer are coupled inducing entanglement. These two descriptions are related to two different subgroups of the $SU(4)$ group of the general four-level system. A recent study performed an analogy between entangled two-spin systems in quantum mechanics and spacetime wormholes in gravity\cite{erdmenger}. The geometric phase was used to label similar entangled states in two-spin and wormholes systems. This underscores the role of the geometric phase in quantum gravity.

Interferometers for gravity tests are likely years away from realization. Here, for the purpose of demonstration, I took $m=10^{-13}$ \, kg, which yielded sufficient phase accumulation in approximately 1\,s in order to reach the phase jump. However, these large masses are out of reach at present time, the largest mass used for interferometry is 10 orders of magnitude smaller $m \sim 10^{-23}$\,kg \cite{fein}. Smaller masses can be used by reducing $d$. However, at small values of $d$, the Casimir-Polder (CP) interaction becomes dominant \cite{schut}. The CP interaction can be reduced by electromagnetic screening \cite{schut} or using magneto-optical effects \cite{cysne}. It can also be reduced by coating  the nanoparticles with a near-unity refractive index material. Indices of refraction as low as $n=1.02$ have been reported for silica and $Al_2O_3$ films \cite{miranda-munoz}. As the CP interaction between two spheres of radius $R$ separated by $d$ is $V_{CP} \sim -\frac{23\hbar c}{4\pi}\frac{R^6}{d^7} (\frac{\epsilon -1}{\epsilon+2})^2$ \cite{schut}, assuming a small nanodiamond core within a coating with a low-index thin film with $\epsilon=1.04$, the ratio of the magnitude of the CP potential I find,
$V_{CP}(\epsilon=1.04)/V_{CP}(\epsilon=5.1) \sim 5.3 \times 10^{-4}$. The material with the lowest index reported is Silanized cellulose for which $\epsilon=1.005$, the CP potential would be smaller by the factor $8.4 \times 10^{-6}$, but it is not a thin film. 
In addition to CP interactions, other spurious interactions and noise can inhibit the observation of interferometric signals are still being minimized, such as black-body radiation \cite{henkel}, phonons \cite{henkel}, rotational effects \cite{japha}, and other sources of noise \cite{toros}. 

The sharp phase jump predicted in the semiclassical gravity can be exploited by preparing the interferometer working point in its vicinity. This is performed in two steps. In the first step, calibration is performed by considering a single spin-one-half interferometer that interacts with a spinless mass. Immediately after the RF pulse, a magnetic gradient pulse is applied for a short time $\ll T$. This pulse brings the phase difference $\phi$ in the vicinity of the singularity $\phi=\phi_c-\delta\phi_0/2$, where $\phi_c=\pi$ is the phase difference at the singularity. Subsequently, the phase difference $\delta \phi_0$ induced by gravitational interaction with the spinless mass drives the interferometer across the jump. After this calibration, the experiment is then performed with two spin one-half interferometers.  If the phase reference   is changed from $0$  to $\delta \phi=\phi-\pi$ in the semiclassical case and to $\delta \phi=(\phi_1+\phi_2)/2-\pi$ in the quantum case, the interferometer phase near the jump is given by,

\begin{equation}
\Phi_C=\arctan \frac{\sin \delta \phi}{\cos \delta\phi-1}, \Phi_Q=\arctan \frac{\xi\sin \delta \phi}{\xi\cos \delta\phi-1}.
\label{eq:cq}
\end{equation}

 In Fig.\,\ref{fig:phase-wpoint}, I show that  semiclassical and quantum gravity interferometric phases can be distinguished even by generating small phases differences around the working point. Theoretically, owing to the sharpness of the jump, a very small $\delta \phi_0$ is required.  However, this advantage is limited by the interferometer phase sensitivity and visibility because the jump occurs in the low-visibility region. For instance, when  $\delta \phi_0= \pi/10$\,rad, as shown in  Fig.\ref{fig:phase-wpoint}A, the semiclassical and quantum have distinct observable signatures. However, $\delta \phi_0=\pi/100$\,rad in Fig.\,\ref{fig:phase-wpoint}B, is almost indistinguishable. However, the visibility is low, it is 15\% and 2\% respectively.  I estimated that by implementing the scheme of Ref.\cite{schut}, adding coating, by shifting the origin near the jump and assuming an interferometer phase sensitivity of $10^{-4}$\,rad, masses around $m \sim 10^{-16}$\,kg would be enough to generate phase differences allowing to distinguish between semiclassical and quantum gravity. 

\begin{figure*}[tbhp]
\centering
\includegraphics[width=\textwidth]{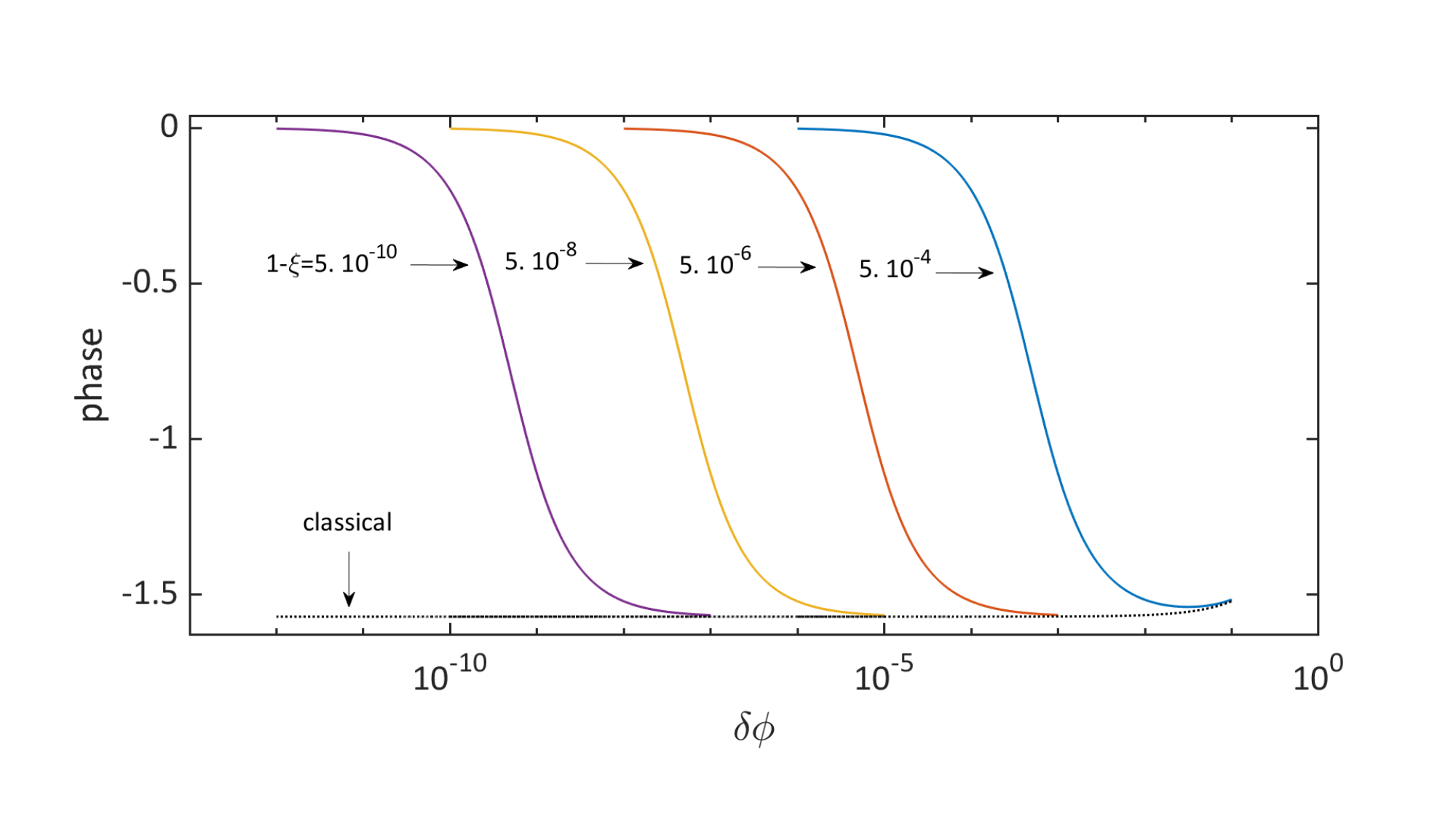}
\vspace{-1cm}
\caption{ The Pancharatnam phase in the dual SGI experiment proposed in Ref.\,\cite{bose} near the semiclassical singularity shown in Fig.\,\ref{fig:phase-vis}. The system is brought at $\delta \phi_0$ from the singularity by a magnetic gradient pulse and then let evolve under the gravitational interaction. Here we use Eq.\,\ref{eq:cq} with values of $\xi$ near to the classical limit $\xi_c=1$ to illustrate the occurence of a large phase difference quantum gravity while $\delta \phi$ is varied in a narrow range near the jump point. The semiclassical phase remains nearly constant for the same values of $\delta \phi$.}
\label{fig:phase-wpointzoom}
\end{figure*}

The relation in Eq.\,\ref{eq:cq} between the interferometric phase $\Phi_Q$ and and the phase difference between the two interferometers $\delta \phi$ near the working point $\phi=\pi$ is further explored in Fig.\,\ref{fig:phase-wpointzoom}. A large phase jump occurs for very small variations of $\delta \phi$. Even very close to the classical limit $\xi \rightarrow 1$, the quantum and classical phases are significantly different. It is clear that even with these optimizations, the realization of such an experiment remains a serious challenge because the signal will be limited by low visibility. Nevertheless, owing to the large variation of $\Phi_Q$ for small $\delta \phi$, the quantum and classical phases can still be distinguished even with $1\%$ visibility.   

The recent controversy surrounding the work of Aziz and Howl \cite{aziz-howl} further underscores the importance of the present approach. Their claim that classical gravity could generate entanglement through virtual-matter processes, if correct, would have severely weakened the interpretational basis of entanglement-based witnesses of quantum gravity. Although this claim has been refuted on theoretical grounds by Di\'osi and others \cite{li-mondal,movw, sienicki, linnemann}, the debate itself reveals a structural vulnerability in proposals that rely solely on the direct detection of spin entanglement as a discriminator: the strength of the entanglement signal is a quantitative, continuous variable, and its interpretation depends sensitively on the theoretical framework assumed. The Pancharatnam phase approach proposed here is immune to this class of objection. The distinction it provides is qualitative and topological in nature - a discontinuous jump versus a smooth inflection point - and this difference is directly tied to the symmetry group governing the evolution of the system, SU(2) in the semiclassical case versus a higher subgroup of SU(4) in the quantum case. No perturbative misidentification of the gravitational source can interpolate continuously between these two topologically distinct behaviors. The Pancharatnam phase witness therefore provides a logically stronger test of the quantum nature of gravity than entanglement strength alone, and its robustness to low visibility makes it experimentally compelling even in the regime where entanglement-based measurements would be unreliable.

\appendix

\section{calculation of the Pancharatnam phase}

 Here we use the kinetic theory developed in Ref.\,\cite{mukunda} for the calculation of the Pancharatnam phase. During the evolution the system wave from $0$ to $t$ the Pancharatnam phase is given by, 

\begin{equation}
\Phi(t)=Arg\left[\langle \Psi(0) | \Psi(t) \rangle \right],
\end{equation}

\noindent where $|\Psi(0)\rangle$ is,  

\begin{equation}
\begin{split}
    |\Psi(0)\rangle &= \frac{1}{\sqrt{2}}(|\uparrow \rangle + |\downarrow \rangle) \otimes \frac{1}{\sqrt{2}}(|\uparrow \rangle + |\downarrow \rangle)\\ 
    &= \frac{1}{2}\left(|\uparrow \uparrow \rangle + |\uparrow \downarrow\rangle + | \downarrow \uparrow \rangle + | \downarrow \downarrow \rangle \right),
\end{split}
\label{eq:phi0}    
\end{equation}

\noindent and $|\Psi(t)\rangle$ is obtained as in Ref.\cite{bose} by assuming that each wavepacket in the superpositions is a source of Newtonian gravitational field. The phase accumulation, at $t$, during the evolution is due to the mutual interaction ($\phi_{ij} \sim G m_im_jt/(\hbar d_{ij})$ for two wavepackets of masses $m_i$, $m_j$ at a distance $d_{ij}$) between the different branches of the interferometers in Fig.\,\ref{fig:sketch}A. Omitting the global phase factor, $|\Psi(t)\rangle$ is given by, 

\begin{equation}
    |\Psi(t)\rangle=\frac{1}{2} \left(|\uparrow \uparrow \rangle + e^{i\phi_1}|\uparrow \downarrow\rangle + e^{i\phi_2}| \downarrow \uparrow \rangle + | \downarrow \downarrow \rangle \right).
 \label{eq:phit}
\end{equation}

For the quantum case I obtain $\Phi_Q$,

\begin{equation}
\Phi_Q=\arctan\frac{\sin \frac{\phi_1+\phi_2}{2} \cos\frac{\phi_1-\phi_2}{2}}{1+\cos\frac{\phi_1+\phi_2}{2} \cos\frac{\phi_1-\phi_2}{2}},
\end{equation}

\noindent where, 

\begin{equation}
\phi_1=\frac{\alpha \Delta x}{d(d-\Delta x)},\,\, \phi_2=-\frac{\alpha \Delta x}{d(d+\Delta x)}
\end{equation}

\begin{equation}
\alpha=\frac{G m_0^2 t}{\hbar},\, m=2m_0.
\end{equation}

For the semiclassical case, the calculation of the Pancharatnam phase $\Phi_c$ is identical to that of a single interferometer, as in Ref.\,\cite{GeoPhaseJump}. I assume that the gravitational field on one interferometer is created by a single mass density on the second interferometer, which is equal to the average mass density of the left and right branches. For equal weights on the branches, this mass is located the center of the interferometer as illustrated in Fig.\,\ref{fig:sketch}B, and at distances   $d-\Delta x/2$ and $d+\Delta x/2$ from the left and right branches of the first interferometer, respectively. I obtain,   

\begin{equation}
\Phi_C=\arctan \frac{\sin \phi}{1+\cos \phi}=\arctan \tan \frac{\phi}{2},
\end{equation}

\noindent where,

\begin{equation}
\phi=\frac{\alpha \Delta x}{(d-\frac{\Delta x}{2})(d+\frac{\Delta x}{2})}.
\end{equation}

\section{Calculation of the visibility}

 The visibility is defined as, $V=\left|\langle \Psi(0) | \Psi(t) \rangle \right|$, using the Eq.\,\ref{eq:phi0},\,\ref{eq:phit} above I find in the quantum case,
 
\begin{equation}
\begin{split}
  V_Q=\frac{1}{2} &\left[1+2\cos(\frac{\phi_1-\phi_2}{2})\cos(\frac{\phi_1+\phi_2}{2}) \right.\\
     &+ \left. \cos^2(\frac{\phi_1-\phi_2}{2})\right]^{\frac{1}{2}},
\end{split}   
 \end{equation}
and in the semiclassical case, 

 \begin{equation}
     V_C=\left| \cos(\frac{\phi}{2}) \right|.
 \end{equation}
 
\begin{acknowledgments}
The author is grateful to Alexandre Bronstein, Sugato Bose, Anupam Mazumdar, and Ron Folman for useful exchanges. This work was funded in part by the Israel Science Foundation Grants No. 856/18 and No. 1314/19. 

The argument concerning the topological robustness of the Pancharatnam phase witness relative to direct spin entanglement-strength witnesses was developed in dialogue with Claude (Anthropic).

\end{acknowledgments}

\end{document}